\documentclass{article}
\pdfoutput=1
\usepackage{geometry}
\usepackage[utf8]{inputenc}
\usepackage{graphicx}
\usepackage{lipsum}
\usepackage{float}
\usepackage{amsmath}
\usepackage{xcolor}
\usepackage{cite}
\usepackage{array}   
\usepackage{diagbox}  
\newcolumntype{P}[1]{>{\centering\arraybackslash}p{#1}}

\title{Ensemble of Pre-Trained Neural Networks for Segmentation and Quality Detection of Transmission Electron Microscopy Images \\ \vspace{0.2in} \small{This manuscript was accepted for publication as part of the Workshop on Machine Learning for Materials Science in conjunction with 28th ACM SIGKDD Conference on Knowledge Discovery and Data Mining. The workshop version can be found here: https://demos.rni.tcsapps.com/mlms-2022/assets/pdfs/10-CameraReady-MLMS\_2022\_Submission\_final.pdf }}

\author{Arun Baskaran $^{1,2}$ \and
        Yulin Lin $^1$ \and
        Jianguo Wen $^1$ \and
        Maria K.Y. Chan $^1$ \and \\
}

\date{%
    $^1$Center for Nanoscale Materials, Argonne National Laboratory, Lemont, IL, USA\\%
    $^2$Currently at Corning Inc., Corning, NY, USA\\%
}

\begin{document}

\maketitle 

\begin{abstract}
Automated analysis of electron microscopy datasets poses multiple challenges, such as limitation in the size of the training dataset, variation in data distribution induced by variation in sample quality and experiment conditions, etc. It is crucial for the trained model to continue to provide acceptable segmentation/classification performance on new data, and quantify the uncertainty associated with its predictions. Among the broad applications of machine learning, various approaches have been adopted to quantify uncertainty, such as Bayesian modeling, Monte Carlo dropout, ensembles, etc. With the aim of addressing the challenges specific to the data domain of electron microscopy, two different types of ensembles of pre-trained neural networks were implemented in this work. The ensembles performed semantic segmentation of ice crystal within a two-phase (ice + water) mixture, thereby tracking its phase transformation to water. The first ensemble (EA) is composed of U-net style networks having different underlying architectures, whereas the second series of ensembles (ER-i) are composed of randomly initialized U-net style networks, wherein each base learner has the same underlying architecture 'i' (i = VGG19, Resnet18, etc.). The encoders of each base learner in both EA and ER were pre-trained on the Imagenet dataset and fine-tuned on the current dataset. The performance of EA and ER were evaluated on three different metrics: accuracy, calibration, and uncertainty. It is seen that EA exhibits a greater pixel classification accuracy and is better calibrated, as compared to ER. While the uncertainty quantification of these two types of ensembles are comparable, the uncertainty scores exhibited by ER were found to be dependent on the specific architecture of its base member ('i') and not consistently better than EA. Thus, the challenges posed for the analysis of electron microscopy datasets appear to be better addressed by an ensemble design like EA, as compared to an ensemble design like ER.  
\end{abstract}


\maketitle

\section{Introduction}

Machine learning (ML) has emerged as a popular tool for performing analysis of image datasets generated from various microscopy techniques \cite{Kalinin2022, Ede2021}. The initial thrust of ML in this field focused on demonstrating the utility of ML to fulfil analysis objectives such as microstructure classification \cite{Kaufmann2020}, structural feature segmentation \cite{Akers2021}, creation of artificial microstructure images to augment training datasets \cite{Ma2020}, etc. A recently emerged focus area is the need to carry forward this body of knowledge and build ML applications that can analyze experimental data as it is produced and with reduced human intervention. An autonomous system presents multiple challenges, especially in the context of electron microscopy datasets. One of them is the limited number of images and labels available for training a ML model. The appropriate training dataset size scales with the number of model parameters, and hence it is generally difficult to train a deep neural network from scratch using only images from the microscopy datasets. A common approach to overcome this challenge is to implement transfer learning \cite{Zhuang2021}. This was followed in the current work. A second challenge in this field is that the image data distribution is dependent on factors such as experimental conditions (temperature, ambient atmosphere, etc.), sample quality, and imaging conditions (focus, tilt, etc.). This makes it possible for shifts to occur from the data distribution in one experiment to the distribution in the next. Keeping these challenges in mind, a few overarching goals were identified for the ML model to achieve. First, the model should continue to provide good performance on new data. Second, the model should be calibrated well within the domain of the data distribution, following which the model should be able to estimate the uncertainty associated with its output. This will aid in the identification of data with excessive noise and data with a shift in distribution from that of the training data. 
Common uncertainty quantification (UQ) approaches incorporated in ML models include the use of Bayesian optimization for parameter estimation, Monte-Carlo dropout, and ensemble learning. A comprehensive review of such techniques can be found in \cite{Abdar2021} and \cite{Gawlikowski2021}. An overview of UQ in a related field to microscopy analysis, biomedical image segmentation, can be found in \cite{Ghoshal2021}. 

\begin{figure}
    \centering
    \includegraphics[width = \linewidth]{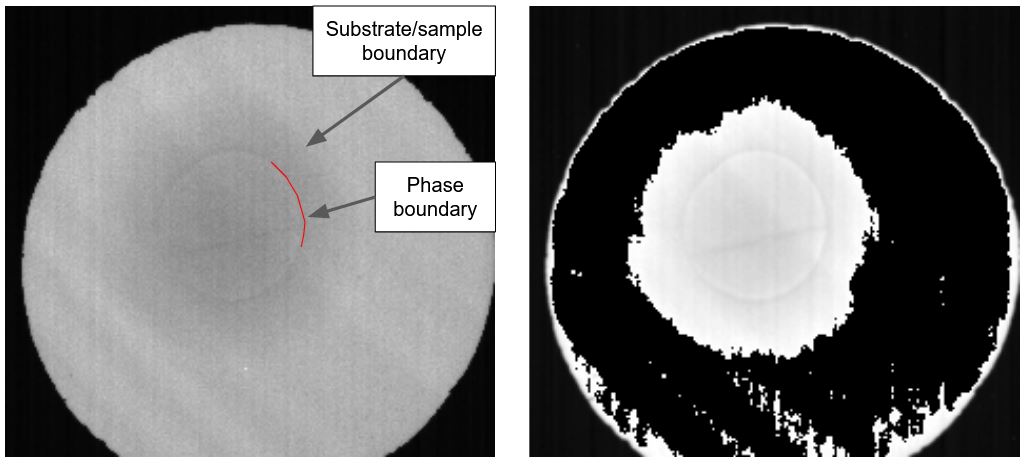}
    \caption{Characteristic examples of (a) a composite image from a dataset generated using the transmission electron microscope and (b) an input to the neural network ensemble. The boundaries separating the ice phase from water and the two phase region from the substrate are annotated in (a). The machine learning ensemble(s) is trained to detect the former. The input (b) is obtained from (a) using pixel intensity thresholding.}
    \label{fig:dataset-summary}
\end{figure}

An important factor behind the increasing use of ML in fields such as electron microscopy has been the development and access to open-source software. For example, software libraries such as tensorflowhub (https://www.tensorflow.org/hub) and segmentation\_models \cite{seg-models2019}, that provide access to pre-trained models, have aided in the application of ML models in fields where large curated training datasets are hard to find. In such cases, it becomes necessary to reuse models that have been pre-trained on large datasets such as the Imagenet \cite{Deng2009}. This is the case with electron microscopy, where generating good quality image data is expensive, time-consuming, and requires a high degree of technical skill. In addition, the creation of manually annotated ground truth data requires additional time from human experts. Thus, reusing pre-trained models can help in making the implementation of ML models more efficient. Examples of transfer learning in the field of ML for microscopy analysis include \cite{Luo2021} and \cite{Matson2019}.

Transfer learning helps in alleviating the problem of limited data. However, it is computationally expensive to quantify uncertainty from single models \cite{Abdar2021}. Thus, a combination of two factors, the necessity for pre-trained networks due to the limited size of manually annotated training data and the ease of implementing UQ using an ensemble approach, motivated our choice of pre-trained ensembles. In this work, we have implemented an ensemble of pre-trained neural networks to implement a workflow for accurate phase segmentation from transmission electron microscopy (TEM) datasets. The specific objectives for the ensemble are as follows: 1) perform accurate segmentation; 2) demonstrate good calibration within the training data domain; and 3) reliably quantify the uncertainty for new data inputs. The use of ensemble of neural networks for ML applications is a relatively well-established field. A review of ensemble learning approaches, in the context of neural networks, can be found in, e.g., \cite{Sagi2018}, \cite{Ganaie2021}, etc. One of the important criteria to consider while designing an ensemble is the diversity among its members. There are multiple ways to incorporate this diversity \cite{Shui2018}, such as by using different starting points through random initialization, different architectures for individual models, different training data subsets, etc. In this work, the utility of two different ensemble designs were tested towards fulfilling the stated objectives. The first ensemble (Ensemble of Architectures, or 'EA') consisted of pre-trained neural network of different model architectures, whereas the second set of ensembles (Ensemble of random initialization of architecture 'i', or ER-i) consisted of pre-trained neural networks of identical architecture 'i'.  

Section 2 outlines the details of the ensemble design, transfer learning, and the training dataset preparation. The major results and discussion with regards to the stated objectives are provided in section 3. 

\begin{figure*}
    \centering
    \includegraphics[width = \linewidth]{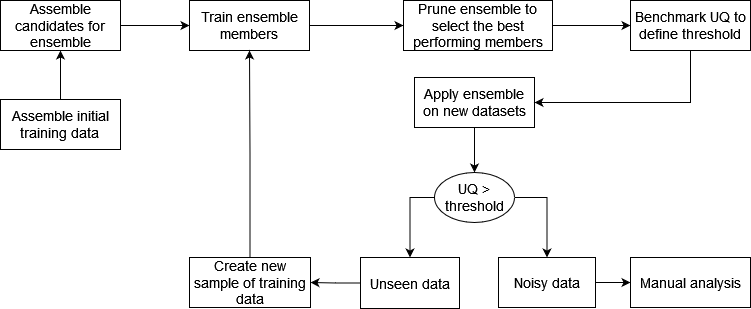}
    \caption{The workflow implemented in this work. The first step is to train the candidates for the base learners for the ensemble in parallel. The trained base learners are added to the ensemble based on a threshold performance on the test set. Then, calibration curve on the test set is extracted from the pruned ensemble and the UQ values are benchmarked. When a new dataset is encountered, the UQ values are measured. If the uncertainty is larger than the benchmark value and the dataset is ascertained to have typical noise levels, samples from the new dataset is added to the training dataset and the ensemble is retrained.  }
    \label{fig:workflow}
\end{figure*}

\section{Methods}

The specific case study chosen for this work is the analysis of phase transformation of nanoscale ice crystals to water. The transformation occurs under a controlled application of temperature, and is recorded in-situ using aberration corrected transmission electron microscopy (ACAT-TEM) \cite{Sinkler2015}. A detailed description about the experimental setup and the scientific motivation for the investigation are provided in a separate manuscript, that is currently in preparation. The instrument produces grayscale images of the sample at a rate between 25-400 frames per second. Figure \ref{fig:dataset-summary}a shows a characteristic image from one of the experiments. This image represents a typical ‘noise level’ (also quantified by signal-to-noise ratio) produced in this experimental setup. Each experiment is run for approximately 25-40 seconds, and the temperature is varied from below melting point ($ T < 0 ^0 C$) to above melting point ($T > 0 ^0 C$). 
This phase transformation occurs at a rate wherein it is appropriate to track the ice crystal at a temporal resolution of 1 s. Thus, composite frames are created by averaging together the frames that are produced for each second. Each experiment produces approximately 25-40 composite images. The two types of boundary present in each image are an outer boundary that separates the ‘water + ice’ region from the background region, and an inner boundary that separates the water phase from the ice phase. A threshold operation based on pixel intensity value is implemented to separate the two-phase region from the background (i.e., the regions separated by the outer boundary), as shown in Figure \ref{fig:dataset-summary}b. Subsequently, the goal of the ML model is to identify the inner boundary and segment the ice phase from water. 

In this case study, five different datasets (D1-D5) were used at various stages of ensemble training. They are summarized in Table \ref{tab:dataset-reference}. Though all these datasets represent the melting or growth of an ice crystal, subtle differences are introduced via factors such as the shape of the initial ice crystal, thermal cycle, length of the experiment, etc. D1 and D2 are the two standard datasets from which samples were drawn for initial training of the ensembles. These datasets consist of images with typical noise levels, as ascertained by the human experts. D3-D5 are datasets that were not included in the initial training process. The purpose of these three datasets was to assess the performance of the trained ensembles on new data that have subtle and uncontrolled differences in the training distribution as compared to the original training dataset. For the first round of training, 10 images each from D1 and D2 were randomly selected. Ground truth is generated by manually annotating each of the 20 images. Geometric augmentation (flipping, rotation, etc.) increased the number of training images and the ground truth labels to 100. The methods involved in the ensemble re-training re-training is briefly discussed in the next section.

\begin{table*}[]
    \centering
    \caption{Datasets and their purpose, with reference to the model training.}
    \begin{tabular}{|c|c|c|}
    \hline
         Dataset ID & No. of composite images & Purpose \\
         \hline
         D1, D2 & Standard dataset & Initial training. \\
         \hline
         D3, D4, D5 & New dataset for inference & Obtaining prediction uncertainty, and model re-training. \\
         \hline
    \end{tabular}
    \label{tab:dataset-reference}
\end{table*}

Two different types of ensembles have been implemented in this study, to compare their performance on the metrics mentioned earlier. Each base member in the ensembles is a pre-trained ‘U-Net’ style neural network. A U-Net \cite{Ronneberger2015} consists of an encoder part and a decoder part. The architecture of the decoder is a mirror image of the encoder, and this architecture is commonly referred to as the backbone. One can obtain different U-net styled models by using different backbones. In the first ensemble implemented in the work (EA), each base learner is a U-net with a different architecture. A candidate pool of pre-trained backbones is assembled for building the ensemble. The complete set of the candidates is provided in \cite{seg-models2019}. A U-net model is initialized using each pre-trained encoder and a decoder with randomly initialized weights. Each of these U-nets are trained in parallel on the training dataset (with a 80\%-20\% split for training-validation). The hyper-parameters for the training process is provided in Table \ref{tab:hyperparameters}. A candidate model is added to the final ensemble if its accuracy score on the validation set, after convergence of the training, is greater than 97\%. In addition, a second set of ensembles were designed by simply randomly initializing different decoder instances of a given model architecture. This set of ensembles is referred to as ER-I, with ‘I’ denoting the backbone for the base model. The different ensembles implemented in this study have been summarized in Table \ref{tab:ensembletypes}.

\begin{table*}[]
    \centering
    \caption{The different ensembles implemented in this study, and their associated purposes. For ER - i, the following backbones are chosen: VGG19, Densenet121, Densenet201, Resnet18, Seresnet50, Inceptionv3}
    \begin{tabular}{|p{10em}|c|c|}
        \hline
         Ensemble name & Ensemble type & Training data \\
         \hline
         EA - T1 & Ensemble formed by U-nets with different architectures & D1 - D2\\
         \hline
         EA - T2 & Ensemble formed by U-nets with different architectures & D1 - D3\\ 
         \hline
         EA - T3 & Ensemble formed by U-nets with different architectures & D1 - D4 \\
         \hline
         EA - T4 & Ensemble formed by U-nets with different architectures & D1 - D2 \\
         \hline
         ER - i \{i = listed in the caption\} & Ensemble formed by randomly initialized U-nets with 'i' as backbone & D1 - D5 \\
         \hline 
    \end{tabular}
    \label{tab:ensembletypes}
\end{table*}

The ensembles are evaluated on multiple metrics. The output at each pixel, class probability obtained from the softmax layer, is collected from all members of the ensemble (Equation \ref{eqn:probagg} ; x denotes a pixel). The average of this set of outputs is considered to be the ensemble's prediction probability at the pixel. The optimum threshold value of probability (0.4) for classification (Equation \ref{eqn:prediction}) is obtained by plotting a Receiver Operating Characteristic (ROC) curve. Pixel classification accuracy (pCA) measures the fraction of pixels that have been correctly classified in an image. Intersection over Union (IoU) measures the overlap between the segmentation map and the ground truth map for the two phases. The uncertainty quantification (UQ) at a pixel is obtained by calculating the standard deviation among the prediction probabilities of the ensemble members (Equation \ref{eqn:UQ}). The average pixel-level UQ values are calculated for each image, and the average image-level UQ values are calculated for each dataset/experiment. The latter value is what is reported in the tables in the following section. 

\begin{equation}
    p_{mean}(x) =  \frac{1}{n} \sum_i p_{\theta_i}(x) ; i = {1,2,...,n}
    \label{eqn:probagg}
\end{equation}

\begin{equation}
    \mathrm{Classification(x)} = 
    \begin{cases}
    1 & p_{mean}(x) \ge 0.4 \\
    0 & p_{mean}(x) < 0.4
    \end{cases}
    \label{eqn:prediction}
\end{equation}

\begin{equation}
    UQ(x) = \sqrt{\frac{\sum_i \big(p_{\theta_i}(x) - p_{mean}(x)\big)^2}{n} }
    \label{eqn:UQ}
\end{equation}

In order to measure uncertainty using this approach, it is important to ensure that the ensemble is well calibrated. The calibration is measured using two different metrics. The Brier calibration score \cite{Rufibach2010} for an input image measures the absolute deviation between the prediction probability at a pixel and its binary label, averaged over the entire image. A perfectly calibrated model has a Brier score of zero. In addition, calibration curves are constructed to assess how well calibrated the model is within the domain of the training data distribution. A calibration curve bins all the pixels in the test dataset according to their prediction probability and plots the fraction of accurately classified pixels in each bin. In a well-calibrated model, this plot follows a x=y relationship.

\begin{table}[]
    \centering
    \caption{Hyper-parameters for the model}
    \begin{tabular}{|c|c|}
    \hline
    Hyper-parameter & Value  \\
    \hline
    Batch size per GPU & 8 \\
    \hline
    Learning rate &  0.001 \\
    \hline
    Training-validation split & 0.2 (20\%) \\
    \hline
    \end{tabular}
    \label{tab:hyperparameters}
\end{table}

The ensemble training and implementation were performed on a hardware setup of 8x NVIDIA-V100 GPUs. The training time until convergence for each base member was approximately 1 hour (between 700-1000 epochs). Re-training a base member until the new convergence point consumed approximately 15 minutes. The inference time for individual images (aggregating predictions from all the models in the ensemble) range from 1-4 seconds per image.

Figure 2 summarizes the overall workflow implemented in the study. The initial training process establishes a benchmark uncertainty on training dataset. After this step, the ensemble is tested on new datasets, and the average uncertainty is calculated for each new dataset. If the average uncertainty is higher than the benchmark value, a decision is made on whether the new dataset has noisy images. If not, then the ensemble is retrained with a few images from the new dataset added to the original training dataset. 

\section{Results \& Discussion}
\begin{figure*}[h]
    \centering
     (a) \hspace{0.45\textwidth} (b) \\
    \includegraphics[width = 0.45\linewidth]{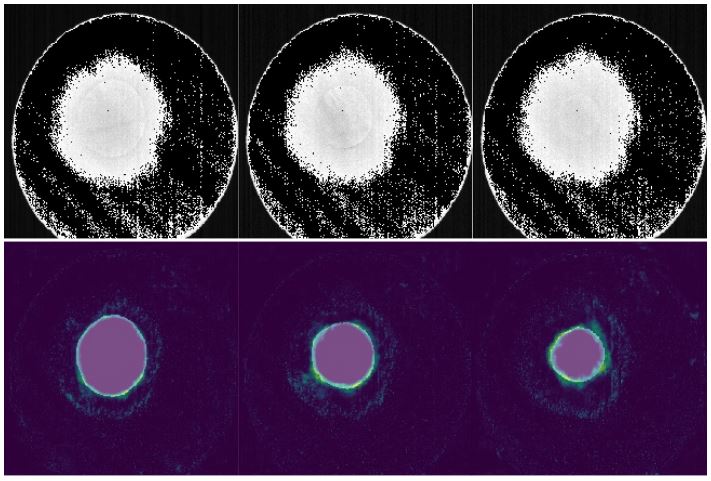}
    \includegraphics[width = 0.45\linewidth]{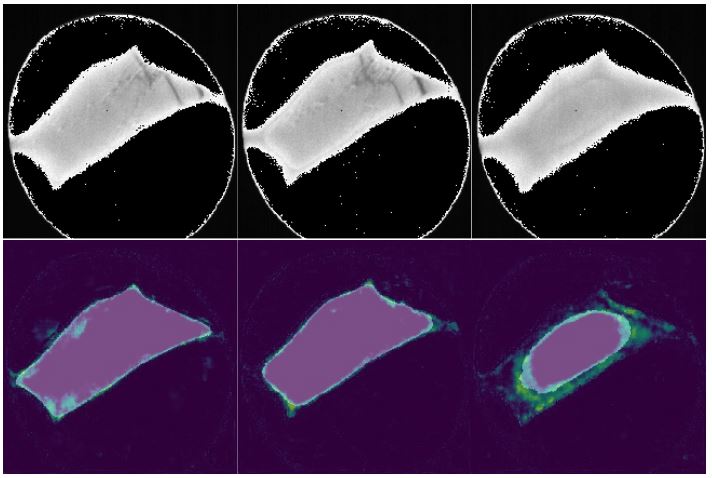}\\
    (c) \\
    \includegraphics[width = 0.45\linewidth]{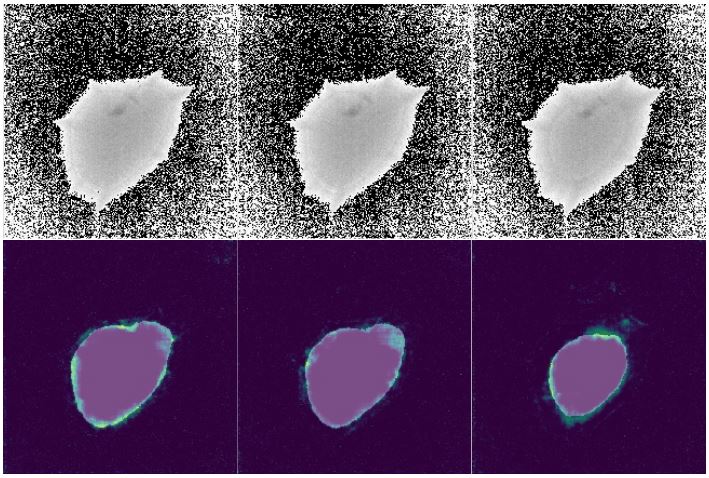}
    \caption{Snapshots of phase segmentation performed by EA for three different datasets used in this study. The three snapshots for each dataset correspond to three different time-steps of the experiment. In the color version, the purple section in the inference denotes the segmentation map whereas the green section denotes the uncertainty values. (a) and (b) correspond to composite images from the two standard datasets D1 and D2, whereas (c) corresponds to the images from the dataset D3.}
    \label{fig:snapshots}
\end{figure*}

The performance of EA (after various stages of training) and ER-I (for different backbone variants) were assessed both qualitatively and quantitatively. To begin with, characteristic segmentation maps obtained using EA-T1 from an image in D2 is shown in Figure \ref{fig:snapshots}a. This is the segmentation obtained after the first round of training.  Table \ref{tab:d1d2stats} lists the pixel classification accuracy (pCA) and the associated uncertainty on test data subset of D1 and D2. The table provides the results from all training cycles of EA as well as a select subset of ER variants. EA, at every training step, outperforms every variant of ER in the pCA evaluation, and thus exhibits better segmentation. The uncertainty obtained by EA-T1 is benchmarked in order to compare it with subsequent datasets. It is also observed, as can be seen from UQ exhibited by the re-trained versions of EA, that the uncertainty for the D1-D2 dataset does not increase significantly after re-training. An UQ estimate is reliable if the ensemble is well-calibrated to the data distribution. The calibration curve exhibited by EA-T1 on the test data from D1 and D2 is shown in figure \ref{fig:calibration}. It can be seen that EA-T1 is well-calibrated for most of the prediction probability range. In comparison, the calibration curves exhibited by two variants of ER (also in figure \ref{fig:calibration}) on the same data domain show that the ER ensemble is relatively poorly calibrated as compared to EA. 

\begin{table*}[]
    \centering
    \caption{Pixel classification accuracy (pCA) and Uncertainty on the two standard benchmark datasets (D1 and D2). For ER - I, i = VGG19, Resnet18, Densenet121, Densenet201}
    \begin{tabular}{|c|c|c|c|c|c|}
    \hline
        & EA – T1 & EA – T2 & EA - T3 & EA - T4 & ER - I  \\
        \hline 
        pCA & 0.996 & 0.986 & 0.975  & 0.974  & 0.926, 0.897, 0.954, 0.958   \\
        \hline
        Uncertainty & 0.015 & 0.016 & 0.012 & 0.012 & 0.012, 0.017, 0.016, 0.014 \\
        \hline
    \end{tabular}
    \label{tab:d1d2stats}
\end{table*}

\begin{table*}[]
    \centering
    \caption{Pixel classification accuracy (pCA) and uncertainty on D3. For ER - I, i = VGG19, Resnet18, Densenet121, Densenet201}
    \begin{tabular}{|c|c|c|c|c|c|}
    \hline
        & EA – T1 & EA – T2 & EA - T3 & EA - T4 & ER - I  \\
        \hline 
        pCA & 0.964  & 0.985  & 0.984  & 0.982  & 0.934, 0.872, 0.954, 0.950 \\
        \hline
        Uncertainty & 0.027 & 0.010 & 0.009 & 0.009 & 0.018, 0.019, 0.024, 0.017  \\
        \hline
    \end{tabular}
    \label{tab:d4stats}
\end{table*}

\begin{table*}[]
    \centering
    \caption{Pixel classification accuracy (pCA) and uncertainty on D4. For ER - I, i = VGG19, Resnet18, Densenet121, Densenet201}
    \begin{tabular}{|c|c|c|c|c|c|}
    \hline
        & EA – T1 & EA – T2 & EA - T3 & EA - T4 & ER - I  \\
        \hline 
        pCA & 0.931  & 0.941  & 0.962 & 0.961  & 0.883, 0.879, 0.923, 0.915  \\
        \hline
        Uncertainty & 0.017 & 0.032 & 0.012 & 0.011 & 0.010, 0.014, 0.017, 0.013  \\
        \hline
    \end{tabular}
    \label{tab:d3stats}
\end{table*}

\begin{figure*}[h]
    \centering
     (a) \hspace{0.25\textwidth} (b) \\
    \includegraphics[width = 0.25\linewidth]{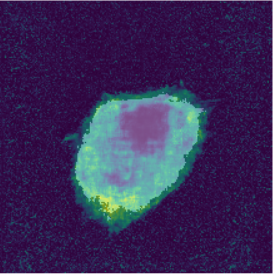}
    \includegraphics[width = 0.25\linewidth]{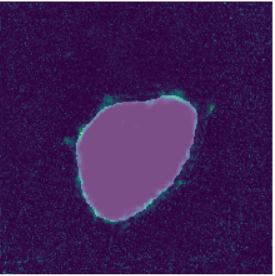}
    \caption{The improvement in prediction uncertainty as a result of the retraining of the ensemble. (a) The prediction uncertainty from the originally trained EA, overlaid on top of its prediction mask. (b) The prediction uncertainty from the re-trained EA, overlaid on top of its prediction mask. It can be observed that the uncertainty is significantly reduced after retraining.}
    \label{fig:std-dev-improvement}
\end{figure*}

\begin{figure*}[h]
    \includegraphics[width = 0.75\linewidth]{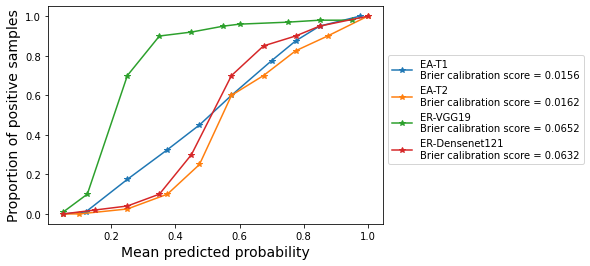}
    \caption{ Calibration curves were constructed for EA-T1, EA-T2, ER-VGG19, and ER-DenseNet121. A calibration curve bins the test set pixels according to the prediction probabilities and calculates the fraction of accurately classified pixels in each bin. A well-calibrated ensemble shows a relationship close to 'y=x'. It is observed here that the EA is well-calibrated after the initial round of training (EA-T1) as well as after re-training with D3 (EA-T2). In comparison, calibration curves from the two ER ensembles are shown here on the D1-D2 datasets. The ER ensembles typically show poor calibration as compared to EA.}
    \label{fig:calibration}
\end{figure*}

The trained ensemble EA-T1 was tested on data from new datasets. The performance of EA-T1 on the data from D3 and D4 is listed in tables \ref{tab:d3stats} and \ref{tab:d4stats}. It was observed that the pCA exhibited by EA-T1 dropped when it encountered new data. In addition, the UQ on the new data showed larger values as compared to the benchmarked values. The increase in the uncertainty implies a shift in the data distribution as compared to the original training data. In order to fine-tune the trained ensemble, two images were randomly selected from each of the new datasets and were manually annotated. These images and labels were added to the original training dataset. The re-trained ensemble (EA-T2 after fine-tuning with D3, EA-T3 after fine-tuning with D4) resulted in an increase in pCA and a concurrent decrease in the associated uncertainty, as observed in tables \ref{tab:d3stats} and \ref{tab:d4stats}. Characteristic segmentation maps obtained using EA-T3 from three images in D3 is shown in figure \ref{fig:snapshots}c. It is also noted that the ensemble remained well calibrated even after the re-training process, as can be seen from the calibration curve exhibited by EA-T2 on the data domain of D1-D3 in figure \ref{fig:calibration}. It is pertinent to highlight two points with regards to the performance of ER variants on D3 and D4. First, EA outperforms the ER variants on the pCA metric, similar to the trend observed for D1 and D2. Thus, EA appears to be better suited to produce segmentation maps compared to ER. Second, when ER encounters data from D4, it exhibits a reduction in uncertainty as compared to the values for D1 and D2, but also exhibits a reduction in the pCA. This suggests that the UQ estimates provided by ER variants are less suited to be used as trigger points for re-training the model. In other words, an ensemble design like EA appears to be better suited to handle adversarial data inputs than an ensemble design like ER.

It is noted that one of the key factors behind designing an ensemble is the diversity among the ensemble members. The preliminary results seen in this work suggests that diversity in model architectures results in a better performing ensemble as compared to diversity through randomized initialization. Further analysis may be required, for example analyzing the final states in the parameter space attained by each ensemble \cite{Fort2020}, to investigate this hypothesis. 

\begin{figure*}[h]
    \centering
    \includegraphics[width = 0.3\linewidth, height = 1.8
    in]{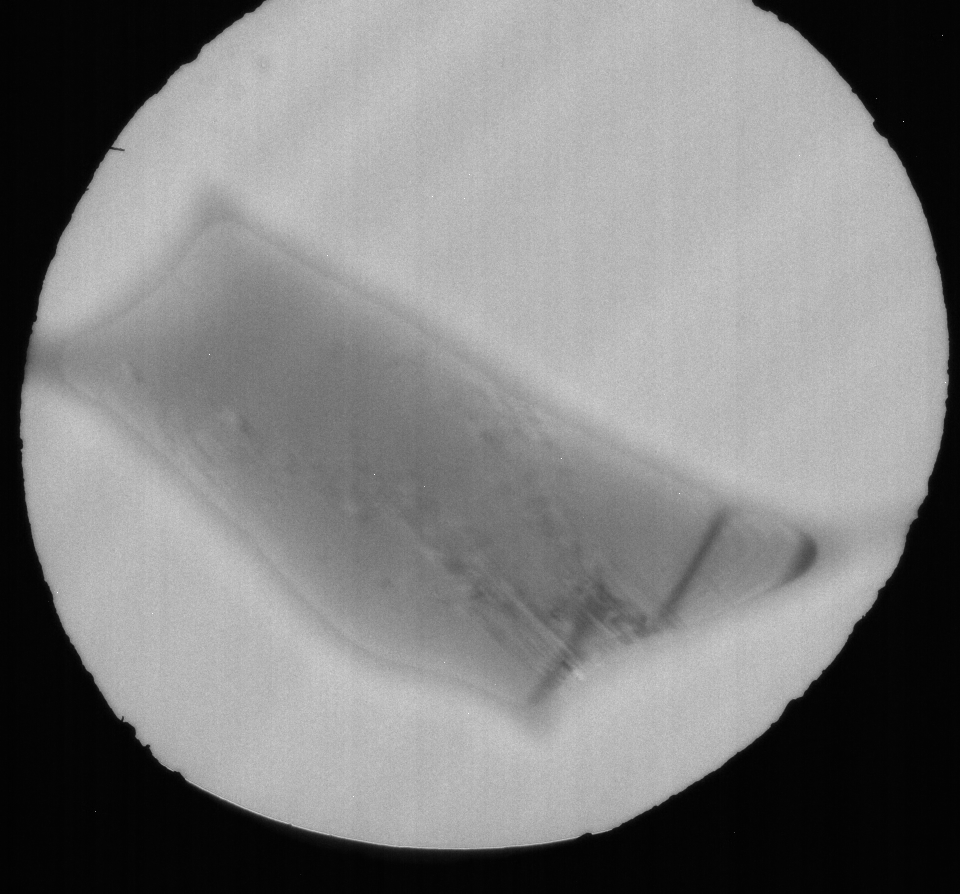}
    \includegraphics[width = 0.32\linewidth, height = 1.8in]{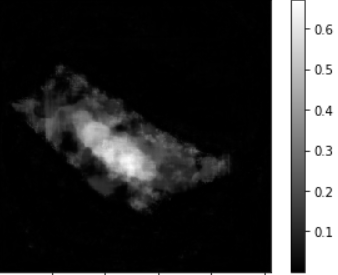}
    \includegraphics[width = 0.32\linewidth, height = 1.8in]{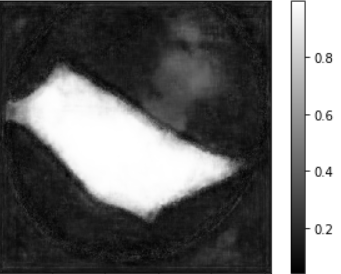}
    \caption{A qualitative assessment of the necessity of transfer learning for this application. The pre-trained ensemble (b) offers a superior segmentation from the two-phase image (a) over the the ensemble that is trained from scratch (c). Despite the simplicity of binary pixel-level classification, the small size of the dataset makes it infeasible to train deep networks from scratch. }
    \label{fig:transferlearning}
\end{figure*}

Finally, a small study was designed to demonstrate the necessity of transfer learning for this problem. Each base member of the EA ensemble was trained from scratch, i.e., the weights of both encoder and decoder were randomly initialized, using the training dataset for EA-T1. The ensemble that was trained from scratch exhibited a test set pCA of 0.843 as compared to the score of 0.996 exhibited by EA-T1. This comparison is supported by a qualitative comparison of the segmentation masks generated by models trained from scratch with the segmentation mask generated by EA-T1, in figure \ref{fig:transferlearning}. Thus, fine-tuning the pre-trained networks is a superior strategy in this low-data regime, despite the simplicity of binary classification. 

Since one of the objectives of the project is to take a step towards integration with an automated analysis platform, it was important to benchmark the time for inference after implementing the ensemble.  As noted in section 2, the training and inference were performed on composite images formed by stacking together the number of frames generated in one second from the instrument. The total time required for frame stacking, image pre-processing, and inference was 3-5 seconds on the hardware setup outlined in the previous section. This feedback time can be improved with better hardware resources as well as designing ensembles specifically geared towards faster inference, for example using lighter models. In the context of building an automated analysis platform, it is important to compare the time required for phase segmentation from trained ensemble with the time taken by manual analysis alone. It takes approximately 3 minutes to manually annotate one image, and hence approximately 75 minutes for manually analyzing a dataset of 25 composite images. It should also be noted that it takes approximately 1 hour to initially train the ensemble (assuming that the base members are trained in parallel). Thus, this framework consumes as much time as manual analysis for the first dataset, but provides a significant reduction in time from the second dataset onwards. Even for a dataset for which the ensemble requires a re-training, there is a considerable reduction in time in re-training the ensemble and extracting inference. 

Another factor to consider in future is the selection of samples from the new dataset for re-training of the ensemble. In this work, two images were randomly selected to be annotated and added to the training dataset. While this approach was sufficient to reduce prediction uncertainties in the current set of datasets, a more systemic approach may be required for cases where there is a larger shift in data distribution. For example, approaches in the field of active learning \cite{Roels2019} can be implemented for selection of new training samples. 

\section{Conclusions}
An ensemble of pre-trained U-net style neural networks is fine-tuned to perform segmentation of ice phase in a two-phase (ice + water) domain from images obtained using a transmission electron microscope. Two different types of ensembles are implemented in this study: an ensemble in which each base learner has a different encoder architecture (EA), and ensembles in which the the base learners have identical architecture but differ simply by the random initialization of the decoders (ER-I ; I denoting the specific encoder architecture). The ensembles were evaluated on three different metrics: segmentation accuracy (pixel classification accuracy), calibration, and uncertainty when facing data from new experiments. It is seen that EA outperforms all variants of ER on the accuracy metrics, resulting in comparatively better segmentation maps. It is also observed that, through the use of calibration curves, EA is better calibrated to the training data domain than the ER variants. This enables a reliable extraction of uncertainty values from new data. Consequently, it is shown that the uncertainty generated by EA on new data can be used as a trigger point to re-train the ensemble on samples from the new dataset. The re-trained ensemble has an improved segmentation performance, lower uncertainty on the data domain, and remains well calibrated. Thus, an  ensemble design like EA appears to be better suited, as compared to an ensemble design like ER, to identify datasets with adversarial inputs such as excessive noise or data distribution shift. A further theoretical investigation is necessary to understand the fundamental cause for the different performances exhibited by these ensembles. 

\section{Availability of code}

The code written during this project for image processing, model training, and evaluation has been made available in the following github repository:\\ https://github.com/MaterialEyes/Vision\_for\_Microscopy. 

\section{Acknowledgments}

Work, including computational work and electron microscopy, performed at the Center for Nanoscale Materials, a U.S. Department of Energy Office of Science User Facility, was supported by the U.S. DOE, Office of Basic Energy Sciences, under Contract No. DE-AC02-06CH11357. This research used resources of the National Energy Research Scientific Computing Center, a DOE Office of Science User Facility supported by the Office of Science of the U.S. Department of Energy under Contract No. DE-AC02-05CH11231.


\begin{thebibliography}{99}
\bibitem{Kalinin2022}Sergei V. Kalinin, Colin Ophus, Paul M. Voyles, Rolf Erni, Demie Kepaptsoglou,
Vincenzo Grillo, Andrew R. Lupini, Mark P. Oxley, Eric Schwenker, Maria K. Y. Chan, Joanne Etheridge, Xiang Li, Grace G. D. Han, Maxim Ziatdinov, Naoya Shibata, and Stephen J. Pennycook. 2022. Machine learning in scanning transmission electron microscopy. Nature Reviews Methods Primers 2, 1 (March 2022), 11. https://doi.org/10.1038/s43586-022-00095-w

\bibitem{Ede2021} Jeffrey M Ede. 2021. Deep learning in electron microscopy. Machine Learning: Science and Technology 2, 1 (mar 2021), 011004. https://doi.org/10.1088/2632-
2153/abd614

\bibitem{Kaufmann2020} Kevin Kaufmann, Chaoyi Zhu, Alexander S. Rosengarten, Daniel Maryanovsky, Tyler J. Harrington, Eduardo Marin, and Kenneth S. Vecchio. 2020. Crystal symmetry determination in electron diffraction using machine learning. Science 367, 6477 (2020), 564–568. https://doi.org/10.1126/science.aay3062 arXiv:https://www.science.org/doi/pdf/10.1126/science.aay3062

\bibitem{Akers2021} Sarah Akers, Elizabeth Kautz, Andrea Trevino-Gavito, Matthew Olszta, Bethany E.Matthews, Le Wang, Yingge Du, and Steven R. Spurgeon. 2021. Rapid and flexible segmentation of electron microscopy data using few-shot machine learning. npj Computational Materials 7, 1 (Nov. 2021), 187. https://doi.org/10.1038/s41524-
021-00652-z

\bibitem{Ma2020} W. Ma, E. J. Kautz, A. Baskaran, A. Chowdhury, V. Joshi, B. Yener, and D. J. Lewis. 2020. Image-driven discriminative and generative machine learning algorithms for establishing microstructure–processing relationships. Journal
of Applied Physics 128, 13 (2020), 134901. https://doi.org/10.1063/5.0013720

\bibitem{Zhuang2021}Fuzhen Zhuang, Zhiyuan Qi, Keyu Duan, Dongbo Xi, Yongchun Zhu, Hengshu Zhu, Hui Xiong, and Qing He. 2021. A Comprehensive Survey on Transfer Learning. Proc. IEEE 109, 1 (2021), 43–76. https://doi.org/10.1109/JPROC.2020. 300455

\bibitem{Abdar2021} Moloud Abdar, Farhad Pourpanah, Sadiq Hussain, Dana Rezazadegan, Li Liu, Mohammad Ghavamzadeh, Paul Fieguth, Xiaochun Cao, Abbas Khosravi, U. Rajendra Acharya, Vladimir Makarenkov, and Saeid Nahavandi. 2021. A review of uncertainty quantification in deep learning: Techniques, applications and challenges. Information Fusion 76 (2021), 243–297. https://doi.org/10.1016/j.inffus.2021.05.008

\bibitem{Gawlikowski2021} Jakob Gawlikowski, Cedrique Rovile Njieutcheu Tassi, Mohsin Ali, Jongseok Lee, Matthias Humt, Jianxiang Feng, Anna Kruspe, Rudolph Triebel, Peter Jung, Ribana Roscher, Muhammad Shahzad, Wen Yang, Richard Bamler, and Xiao Xiang Zhu. 2021. A survey of uncertainty in deep neural networks. arXiv:2107.03342v3 (2021).

\bibitem{Ghoshal2021} Biraja Ghoshal, Allan Tucker, Bal Sanghera, and Wai Lup Wong. 2021. Estimating uncertainty in deep learning for reporting confidence to clinicians in medical image segmentation and diseases detection. Computational Intelligence 37, 2 (2021), 701–734. https://doi.org/10.1111/coin.12411

\bibitem{seg-models2019}Pavel Yakubovskiy. Segmentation models, 2019. https://github.com/qubvel/segmentation\_model

\bibitem{Deng2009}Jia Deng, Wei Dong, Richard Socher, Li-Jia Li, Kai Li, and Li Fei-Fei. 2009. ImageNet: A large-scale hierarchical image database. In 2009 IEEE Conference on Computer Vision and Pattern Recognition. 248–255. https://doi.org/10.1109/CVPR.
2009.5206848

\bibitem{Luo2021} Qixiang Luo, Elizabeth A. Holm, and Chen Wang. 2021. A transfer learning approach for improved classification of carbon nanomaterials from TEM images. Nanoscale Adv. 3 (2021), 206–213. Issue 1. https://doi.org/10.1039/D0NA00634

\bibitem{Matson2019} Thomas Matson, Max Farfel, Nathan Levin, Elizabeth Holm, and Chen Wang. 2019. Machine Learning and Computer Vision for the Classification of Carbon Nanotube and Nanofiber Structures from Transmission Electron Microscopy Data. Microscopy and Microanalysis 25, S2 (2019), 198–199. https://doi.org/10.1017/S1431927619001727

\bibitem{Sagi2018} Omer Sagi and Lior Rokach. 2018. Ensemble learning: A survey. WIREs Data Mining and Knowledge Discovery 8, 4 (2018), e1249. https://doi.org/10.1002/widm. 

\bibitem{Ganaie2021} Mudasir A. Ganaie, Minghui Hu, Mohammad Tanveer, and Ponnuthurai N. Suganthan. 2021. Ensemble deep learning: A review. CoRR abs/2104.02395 (2021). arXiv:2104.02395 https://arxiv.org/abs/2104.02395

\bibitem{Shui2018} Changjian Shui, Azadeh Sadat Mozafari, Jonathan Marek, Ihsen Hedhli, and
Christian Gagné. 2018. Diversity regularization in deep ensembles. https:
//doi.org/10.48550/ARXIV.1802.07881

\bibitem{Sinkler2015} Wharton Sinkler, Sergio I. Sanchez, Steven A. Bradley, Jianguo Wen, Bhoopesh Mishra, Shelly D. Kelly, and Simon R. Bare. 2015. Aberration-Corrected Transmission Electron Microscopy and InSitu XAFS Structural Characterization of Pt/-Al2O3 Nanoparticles. ChemCatChem 7, 22 (2015), 3779–3787. https://doi.org/10.1002/cctc.201500784 

\bibitem{Ronneberger2015} Olaf Ronneberger, Philipp Fischer, and Thomas Brox. 2015. U-Net: Convolutional Networks for Biomedical Image Segmentation. In Medical Image Computing and Computer-Assisted Intervention – MICCAI 2015, Nassir Navab, Joachim Hornegger, William M. Wells, and Alejandro F. Frangi (Eds.). Springer International
Publishing, Cham, 234–241.

\bibitem{Rufibach2010} Kaspar Rufibach. 2010. Use of Brier score to assess binary predictions. Journal of
clinical epidemiology 63, 8 (2010), 938–939.

\bibitem{Fort2020} Stanislav Fort, Huiyi Hu, and Balaji Lakshminarayanan. 2020. Deep ensembles: A loss landscape perspective. arXiv:1912.02757v2 (2020)

\bibitem{Roels2019} Joris Roels and Yvan Saeys. 2019. Cost-efficient segmentation of electron microscopy images using active learning. arXiv:1911.05548



\end{thebibliography}

\end{document}